\documentclass[journal]{vgtc}                     

\onlineid{0}

\vgtccategory{Research}

\vgtcpapertype{please specify}

\title{Occlusion-Free Conformal Lensing for Spatiotemporal \\Visualization in 3D Urban Analytics}

\author{Roberta Mota, Julio D. Silva, Fabio Miranda, Usman Alim, Ehud Sharlin, Nivan Ferreira}

\authorfooter{
\item
 Roberta Mota, Julio D. Silva, Usman Alim, and Ehud Sharlin are with the University of Calgary, Canada. E-mails: \{roberta.cabralmota, jd.silva, ualim, ehud\} @ucalgary.ca.
\item
Fabio Miranda is with the University of Illinois Chicago, United States. E-mail: fabiom@uic.edu.
\item
 Nivan Ferreira is with the Universidade Federal de Pernambuco, Brazil. E-mail: nivan@cin.ufpe.br.
}

\abstract{%
    The visualization of temporal data on urban buildings, such as shadows, noise, and solar potential, plays a critical role in the analysis of dynamic urban phenomena. However, in dense and geographically constrained 3D urban environments, visual representations of time-varying building data often suffer from occlusion and visual clutter.
    To address these two challenges, we introduce an immersive lens visualization that integrates \textit{i}) a view-dependent cutaway de-occlusion technique and \textit{ii}) a temporal display derived from a conformal mapping algorithm.
    The mapping process first partitions irregular building footprints into smaller, sufficiently regular subregions that serve as structural primitives. These subregions are then seamlessly recombined to form a conformal, layered layout for our temporal lens visualization.
    The view-responsive cutaway is inspired by traditional architectural illustrations, preserving the overall layout of the building and its surroundings to maintain users' sense of spatial orientation. This lens design enables the occlusion-free embedding of shape-adaptive temporal displays across building facades \textit{on demand}, supporting rapid time-space association for the discovery, access and interpretation of spatiotemporal urban patterns.
    Guided by domain and design goals, we outline the rationale behind the lens visual and interaction design choices, such as the encoding of time progression and temporal values in the conforming lens image. A controlled user study compares our approach against conventional juxtaposition and x-ray spatiotemporal designs.
    Results validate the usage and utility of our lens, showing that it improves task accuracy and completion time, reduces navigation effort, and increases user confidence. From these findings, we distill design recommendations and promising directions for future research on spatially-embedded lenses in 3D visualization, urban analytics, and related domains.
}

\keywords{Focus+context, lenses, visualization, virtual reality, immersive analytics, urban analytics, spatiotemporal urban data}

\teaser{
  \centering
  \includegraphics[width=\linewidth]{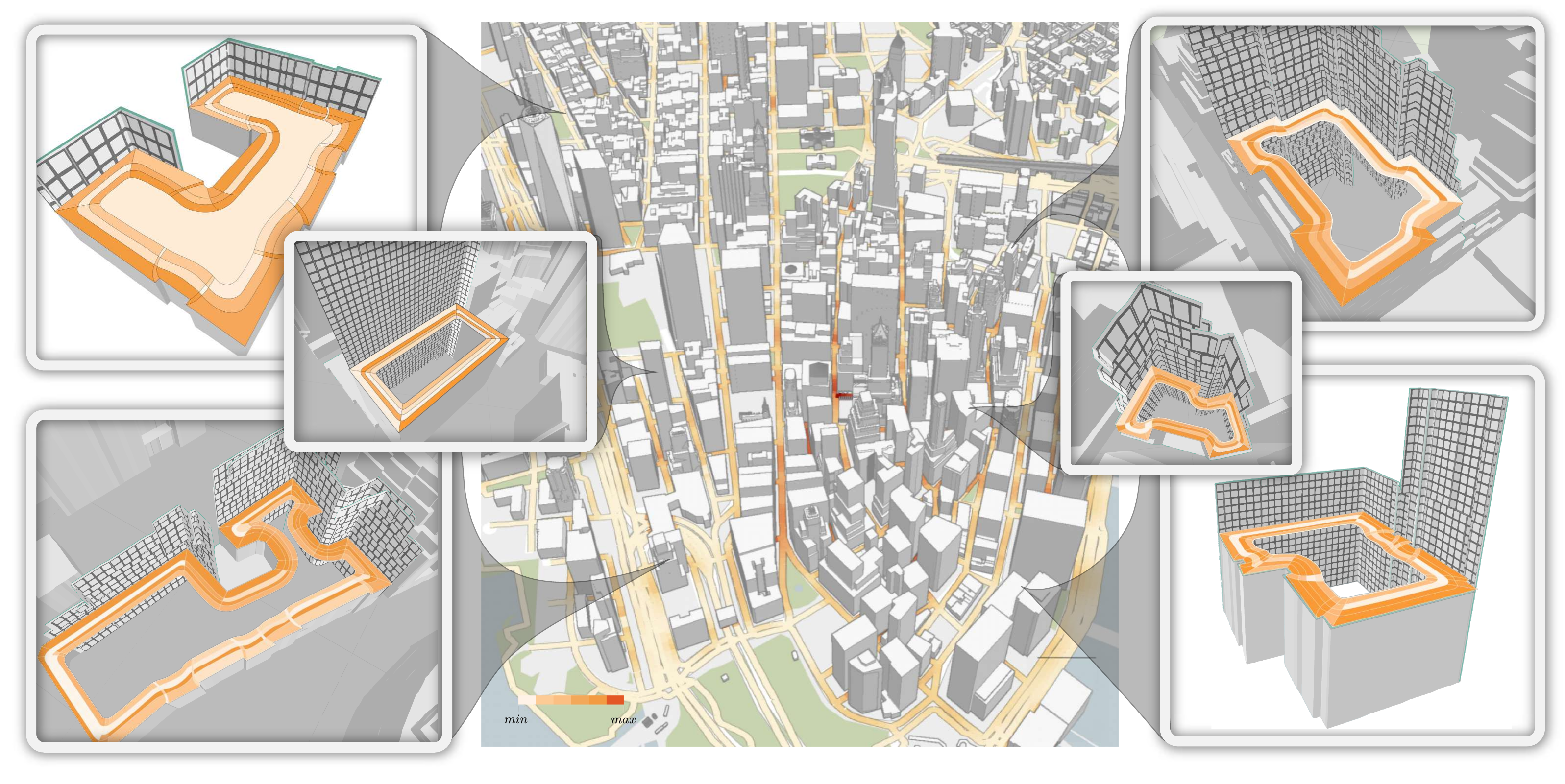}
  \caption{
    Our lens \textit{augments} 3D urban visualizations with temporal displays that seamlessly conform to building footprints, while \textit{suppressing} self-occluding geometry to maintain an unobstructed line of sight to the time-series visualizations in focus.
    While the lens accommodates traditional allocentric analysis, where users examine data from a detached, top-down perspective, it is designed for egocentric analysis, enabling users to explore spatiotemporal data from within immersive 3D urban environments.
  }
  \label{fig:teaser}
}

\graphicspath{{figs/}{figures/}{pictures/}{images/}{./}} 

\usepackage{mathptmx}                  

\usepackage{xcolor}
\usepackage{textcomp}
\usepackage{enumitem}
\usepackage{array}
\usepackage{scalerel}
\usepackage{amsmath}
\usepackage{amssymb}
\usepackage{stfloats}

\definecolor{myGrey}{HTML}{4d4d4d}
\definecolor{meanColor}{HTML}{69b3a2}
\definecolor{pairwiseColor}{HTML}{446df6}
\definecolor{evidenceColor}{HTML}{e6550d}

\definecolor{sanddune}{rgb}{0.59, 0.44, 0.09}
\definecolor{seagreen}{rgb}{0.18, 0.55, 0.34}

\newcommand{\visLabel}[1]{{\fontfamily{cmtt}\selectfont {#1}}}

\newcommand{\hide}[1]{}

\newcommand{\removetext}[1]{\textcolor{red}{\sout{#1}}}
\renewcommand{\removetext}[1]{\unskip}

\newcommand{\highlight}[1]{\textcolor{black}{#1}}

\newcommand{\iconAsterisk}{\raisebox{-0.25ex}{\includegraphics[width=1em]{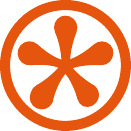}}}

\newcommand{\iconPress}{\raisebox{-0.5ex}{\includegraphics[width=1em]{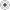}}}

\newcommand{\iconDrag}{\raisebox{-0.5ex}{\includegraphics[width=2em]{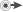}}}

\begin{document}



\maketitle

\section{Introduction}
Recent advances have increased the availability of temporal data on urban buildings, such as shadows, noise, and solar potential~\cite{miranda2024state}. This opens opportunities for novel visualizations of time-varying building data, which play a critical role in the analysis of dynamic urban phenomena.
In particular, immersive visualizations have gained traction in urban visual analysis ~\cite{Chen2017, chen2021urbanrama, wu2017efficient, wagner2024reimagining, chen2017immersive}, as immersive technology allow users to exploit their innate 3D perception and interaction abilities to explore 3D urban data. 
Despite this, in dense and geographically
constrained 3D urban environments, immersive visualizations of time-varying building data often suffer from \textit{visual clutter} and \textit{occlusion}. 
First, when visualizing time-varying data in the building grid geometry, it is challenging to include all information into a single visual image and often leads to \textit{clutter}, excessive visual information that makes the visualization \removetext{too confusing for users to interpret and gain insight from it} hard to interpret.
Second, visualizations must deal with data hidden by the building geometry due to \textit{self-occlusion}, where facades from the building itself occlude each other \removetext{from some viewpoint}.
If facades containing relevant information are partially or fully occluded, analysts must repeatedly shift camera perspectives and mentally retain data across views to construct a holistic mental model during analysis, ultimately hindering their ability to derive data insights.
Together, clutter and occlusion have a detrimental impact on discoverability, accessibility, and interpretability of time-varying building data \removetext{in 3D urban environments}. 

To mitigate these issues, we investigate an immersive visualization based on \textit{lenses}, a focus+context technique that enable users to locally modify the visual representation of the data underlying a focus area of the screen \cite{tominski2017interactive}. 
\removetext{In general, a lens can modify a visualization by \textit{altering} existing content, \textit{suppressing}
irrelevant content, \textit{augmenting} with new content, or even a combination thereof.}
Our lens \textit{augments} the visualization with a temporal plot that conforms to the building footprint, and \textit{suppresses} self-occluding building geometry to maintain an unobstructed line of sight to the temporal display (Fig.~\ref{fig:teaser}).
    By doing so, our lens visualization integrates spatial and temporal data into a single unoccluded image: the overlaid lens image portrays the temporal variation, while the building grid geometry provides spatial context.
    This aims to enable rapid time-space association, and ultimately facilitate the discovery, access, and interpretation of spatiotemporal patterns.
Our contributions can be summarized as follows:

\begin{itemize}[leftmargin=*, nolistsep]
    \item A {\textbf{mapping technique}} to generate conformal maps that adapt to the irregular geometries of building footprints. These maps serve as geometric layouts for our lens visualization.

    \item \highlight{An {\textbf{immersive lens visualization}} to support time-varying building data analysis by mitigating self-occlusion and clutter. 
    To this end, the lens integrates \textit{i}) a view-dependent cutaway for de-occlusion and \textit{ii}) a temporal display derived from the conformal mapping algorithm and embedded within the building.
    } 
    
    \removetext{The lens \textit{augments} the visualization with a temporal plot that conforms to the building footprint, and \textit{suppresses} self-occluding building geometry to maintain an unobstructed line of sight to the temporal display.
    By doing so, the lens visualization integrates spatial and temporal data into a single unoccluded image: the overlaid lens image portrays the temporal variation, while the building grid geometry provides spatial context.
    This aims to enable rapid time-space association, and ultimately facilitate the discovery, access, and interpretation of spatiotemporal patterns.
    }

    \item A \textbf{controlled evaluation} to validate the usefulness of our lens visualization in practical, task-driven urban scenarios. Following, we abstract design recommendations to guide future research.
\end{itemize}


\section{Related Work}
\label{sec:relatedWork}
This section reviews prior work on visual design strategies employed in spatiotemporal data, occlusion management, and lenses, in both conventional and more immersive environments.

\subsection{Visualization of Spatiotemporal Data}
\highlight{Andrienko \textit{et al.}} conducted a survey on information visualizations for spatiotemporal 2D data and identified two visual integration strategies: \textit{linked view} and \textit{\textit{embedded view}} \cite{Andrienko2010}. Chen \textit{et al}. later extended this taxonomy to 3D urban data in immersive spaces \cite{Chen2017}.
\textbf{Linked views} present spatial and temporal data in various separate visual spaces \cite{turkay2014attribute}, and thus impose \removetext{A major limitation of this design is} cognitive overhead due to spatial detachment between views, forcing users to mentally relate temporal patterns to their spatial referents.
\highlight{Additionally, the need for extra screen real estate may be detrimental in immersive environments constrained by limited field-of-view, such as head-mounted displays.}
\textbf{Embedded views}, in contrast, co-locate spatial and temporal data within the same view \cite{andrienko2004interactive, sun2016embedding}. In urban visualizations, spatial context is typically conveyed through the geometry of the city, \textit{e.g.}, building footprints, while overlaid temporal plots portray changes over time. While this design enhances space-time correspondence, it might suffer from clutter and occlusion.

Furthermore, Kim \textit{ et al.}'s taxonomy classifies scientific visualizations for spatiotemporal 3D data into four design patterns: \textit{juxtaposition}, \textit{superimposition}, \textit{interchangeable}, and \textit{explicit encoding} \cite{kim2017comparison}.
\textbf{Juxtaposition} displays multiple temporal instances simultaneously, each rendered in a distinct coordinate space, via side-by-side views \cite{johnson2019bento}.
While this design aids temporal comparison without taxing the user's memory, it introduces spatial-temporal decoupling and a trade-off in screen space allocation, paralleling the limitations of linked views.
\textbf{Interchangeable} renders one temporal instance at a time, typically switching via either user interaction or animation \cite{akiba2009aniviz}. All instances are spatially co-registered, and each is displayed at a full scale. However, users must rely on memory to mentally compare temporal states.
\textbf{Superimposition} displays multiple temporal instances concurrently within the same coordinate space \cite{van2014comparative}. It is the only design that ensures spatial co-registration, simultaneous visibility, and preservation of the original data.
\textbf{Explicit Encoding} forgoes displaying original data and instead derives composite attributes, \textit{e.g.}, the difference between two time steps \cite{tory2001visualization}. This design enables full-scale depiction of derived data but limits interpretability, as individual instances are not visible.
\removetext{Particularly, }A recent survey \removetext{classifies urban visualizations based
on the aforementioned info-vis and sci-vis spatiotemporal designs} \cite{mota2022comparison} on spatiotemporal urban visualization reveals a strong reliance on \textit{interchangeable} designs in 3D scenarios, while \textit{linked views} are more prevalent in 2D visualizations. In contrast, our lens combines both \textit{embedded} and \textit{superimposition} strategies to promote tighter spatial-temporal coupling in 3D urban visualizations.

\begin{figure*}[t!]
    \centering
    \includegraphics[width=1\linewidth,keepaspectratio=true]{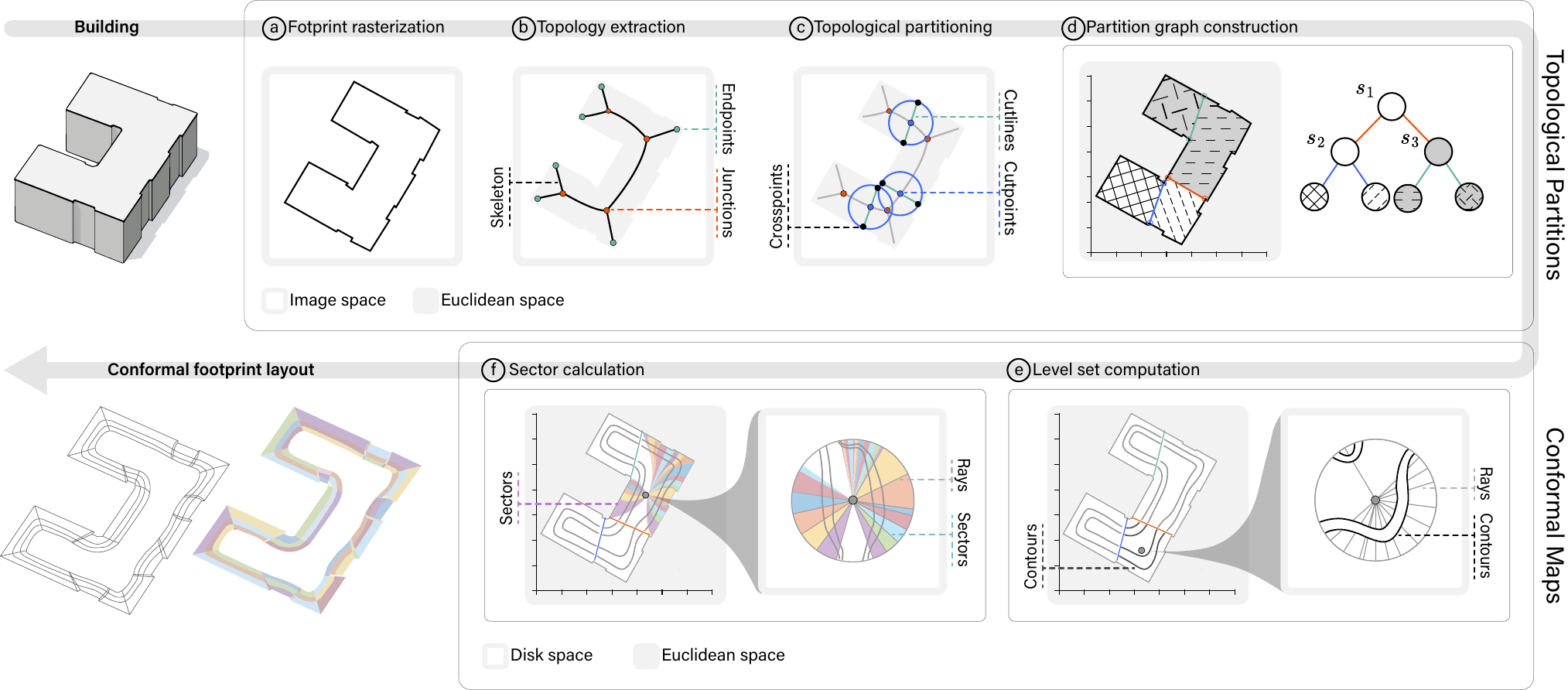}
    \caption{Pipeline illustrating our conformal mapping algorithm. (\textit{Top}) The first stage decomposes the building footprint into smaller parallelogram approximations.
    The recursive decomposition proceeds through successive subdivisions $S_i$, each introducing additional polygonal subregions and corresponding graph nodes, ultimately forming a hierarchical topological graph.
    (\textit{Bottom}) The second stage computes a conformal map for each partition, and subsequently stitches  them together to form a unified geometric layout.}
    \label{fig:mappingAlgorithm}
\end{figure*}

\subsection{Visualization of Occluded 3D Data}
\noindent Elmqvist and Tsigas' taxonomy of de-occlusion strategies comprise five categories: \textit{multiple viewports}, \textit{projection distorter}, \textit{tour planner}, \textit{volumetric probe}, and \textit{x-ray} \cite{elmqvist2008taxonomy}.
\textbf{Multiple viewports} display target objects from different perspectives using two or more separate views, typically, a primary viewport alongside smaller secondary views. While this design offers comprehensive coverage, it demands careful camera placement to eliminate occlusion fully and often increases cognitive load as users must mentally integrate perspectives across views \cite{cockburn2009review, baudisch2002keeping, chen2021urbanrama}. 
\textbf{Projection distorter} merges multiple projection views to reveal as much of occluded targets as possible in a single frame. However, this often results in perceptually-inconsistent images that users find disorienting \cite{wu2017efficient}.
\textbf{Tour planner} involves a predefined or automatically- generated camera path that traverses target locations, presenting them sequentially \cite{elmqvist2007tour}.
\textbf{Volumetric probe} performs object-space deformations in response to spatial probes controlled by users \cite{chen2017immersive, chen2021urbanrama}. Although effective in exposing occluded elements, such deformation may distort surrounding spatial context.
\textcolor{black}{\textbf{Virtual x-ray} \removetext{methods} enhances visibility by rendering occluding geometry semi-transparent or invisible \cite{looser2004through}. However, it can diminish perception of depth and spatial relationships between objects.
Our approach adopts a view-dependent \textit{x-ray} design, to reveal temporal displays embedded within buildings while retaining sufficient spatial context from the surrounding urban geometry.}

\subsection{Lens Visualization for 3D Data}
The shape of a lens can be classified as either \textit{fixed} or \textit{dynamic}. Most lenses have a \textbf{fixed} shape, typically using primitive forms such as circles, disks, or boxes, though there are no strict limitations on more complex geometries \cite{mota20183de, spindler2009paperlens, gasteiger2011, trapp2008}. Fixed-shape lenses offer simplicity and familiarity, which can lower the learning curve. However, they may fail to conform to irregular data structures, leading to poor coverage of relevant features, inclusion of irrelevant ones, and increased cognitive load as the user must manually disambiguate them.
In contrast, some lens techniques employ dynamic shapes, which adapt by either the user or underlying data. In \textbf{user-driven dynamic} lenses, the shape is commonly specified through interactions such as sketching, brushing, or lassoing \cite{kister2015bodylenses, rocha2018decal}. These enable greater precision and flexibility in capturing irregular features, better reflecting user intent. However, this comes at the cost of increased interaction effort, \textit{e.g.}, manual drawing can become cumbersome in dense 3D spaces.
On the other hand, \textbf{data-driven dynamic} lenses automatically adapt their shape to align with data features such as boundaries, clusters, or paths \cite{pindat2012, ion2013}. This reduces user effort, facilitates automated feature discovery, and may improve interpretability by ensuring the lens conforms meaningfully to the data. Nonetheless, their behavior may become unpredictable in noisy \removetext{or ambiguous} data, likely making it difficult for users to understand the auto-generated shape.
Our lens employs data-driven automation to adapt its shape to the building footprint, ensuring strong spatial coherence between the temporal lens image and the underlying spatial geometry.
\section{Conformal Mapping Algorithm}
\label{sec:mappingAlgorithm}

\noindent \textbf{Problem definition.} A major challenge in embedding temporal displays onto buildings is the mismatch between the rigid, regular shapes typically used in lenses, \textit{e.g.}, circles \removetext{, squares,} or disks, and the often irregular geometry of building footprints.
To address this, our \removetext{core} approach segments the footprint into smaller, sufficiently regular regions that serve as structural primitives of our lens visualization. These segments are then seamlessly recombined to form a conformal, integrated visual layout \removetext{as illustrated in} (Fig. \ref{fig:mappingAlgorithm}).
\highlight{The remainder of this section presents an overview of the algorithm, with further details provided in the supplemental materials.}

\subsection{Topological Signature Partitioning}

The first stage of the mapping algorithm is heuristic in nature and grounded in the observation that building footprints are often composed of parallelepipeds. \removetext{That is, in two dimensions, these footprints are typically manifest as a collage of parallelograms. Therefore, }
The algorithm begins by identifying the topological signatures of the parallelograms comprising a given footprint, and segmenting them as a \textit{directed graph}. To ensure algorithm termination, cycles in the building footprint are removed to produce an acyclic graph. 
The main steps of this initial stage, each denoted as $S_i$,  are as follows:

\vspace{1mm}
\noindent \textbf{S$_1$.} \textit{Topological signature extraction.} To construct the topological signatures, we begin by rasterizing the footprint polygon into a regular grid image.
Next, we extract the {skeleton} of the binarized image $S_b$, along with its corresponding {junctions} and {endpoints} \highlight{(Fig. \ref{fig:mappingAlgorithm}--\textit{a,b})}. \removetext{Junctions
are detected by intersecting vertical and horizontal components in
the binarized skeleton image $S_b$, using morphological operations:}

\removetext{
\vspace{-2mm}
\begin{equation}
    J = \text{centroids}\left( (S_b \circ H) \cap (S_b \circ V) \right) 
\end{equation}
\vspace{-4mm}
}

\removetext{\noindent where \( H \) and \( V \) denote horizontal and vertical structuring, respectively, \( \circ \) represents the morphological opening operator, and \( \cap \) represents the pixel-wise intersection between the two resulting binary images. To compute endpoints, we apply a convolution kernel \( H \in \mathbb{Z}^{3 \times 3} \) over \( S_b \). Each pixel in the resulting image is defined by:}

\removetext{
\vspace{-2mm}
\begin{equation}
    C(x, y) = (S_b * H)(x, y)    
\end{equation}
\vspace{-4mm}
}

\removetext{\noindent where \( C(x, y) \) encodes the local pixel neighborhood, \textit{i.e.}, the number of white 8-connected neighbors in \( S_b \). Next, we extract a set of candidate points \( P = \{(x_i, y_i)\} \) by selecting pixels in \( C \) whose values correspond to three or more white neighbors, which are indicative of branching or terminal points in the skeleton. Finally, endpoints are determined via proximity-based classification: a candidate point \( p \in P \) is an endpoint if it lies beyond a distance threshold from all detected junction points.}

\vspace{1mm}
\noindent \textbf{S$_2$.} \textit{Topological partitioning.} Given the skeleton image along with its junctions and endpoints, we proceed to decompose the footprint into parallelograms, following our initial heuristic assumption. To this end, we note that a parallelogram exhibits a simple topological signature when analyzed via its medial axis: it consists of two trifurcation junctions connected by a central segment, with each junction branching into two endpoints, effectively forming a pair of connected ``Y'' shapes. To segment the footprint into constituent parallelograms, we treat each trifurcation point from the junction set \removetext{$J$} as a seed. For each seed pair, we identify the white skeleton segment in $S_b$ that connects them. Along this segment, we search for the pixel whose inscribed circle intersects two opposing sides of the polygon and has the smallest radius. We refer to this pixel as a \textit{cutpoint}, and the segment that connects the two opposing intersections through this medial point as a  \textit{cutline} \highlight{(Fig. \ref{fig:mappingAlgorithm}--\textit{c})}.

\vspace{1mm}
\noindent \textbf{S$_3$.} \textit{Partition graph construction.} Given the medial cutlines, we segment the original footprint polygon into a set of non-overlapping subregions, each approximating a parallelogram under the medial-axis-based decomposition heuristic. These are assembled into a directed graph $G=(V, E)$, where each node \( v \in V \) refers to a subpolygon annotated with its corresponding topological signature, and each directed edge \( e \in E \) represents the geometric adjacency \highlight{(Fig. \ref{fig:mappingAlgorithm}--\textit{d}).}


\subsection{Conformal Maps, Contours \& Sectors}

\textcolor{black}{The second stage of the mapping algorithm computes conformal maps \removetext{from the unit disks onto the corresponding} for the \textit{graph nodes} \removetext{, \textit{i.e.} the footprint subregions, aiming to ensure that the resulting
mappings are as well-behaved as possible, i.e., as close to isometric as possible to minimize distortion and avoid introducing artifacts into the visualization. To achieve this, we employ }
, using the Schwarz–Christoffel \highlight{(SC)} mapping \removetext{which are conformal by construction
and amenable to be computed efficiently} \cite{driscoll1996algorithm}.
A conformal map comprises two components: \textit{i}) \textit{level sets} \removetext{of the distance function, \textit{i.e.}, the set of}, each defined as points equidistant from the footprint boundary, and \textit{ii}) \textit{sectors} defined by curves that partition the level sets into distinct facades of the building footprint. The main steps in this stage are:}

\vspace{1mm}
\noindent \highlight{\textbf{S$_4$.} \textit{Level set computation.} Let $d : W_k \rightarrow \mathbb{R}$ denote the distance from a point in the footprint subregion $W_k$ to its nearest boundary point, and let $\phi_k : D_k \rightarrow W_k$ be a SC conformal map from the unit disk $D_k$. We compute level sets by pulling back the distance function to the disk,  $d_k(z) = d(\phi_k(z))$ where \(z \in D_k\). A contour at a target distance $d_W$ is then constructed by \textit{i}) \removetext{sampling radial directions \( \theta_i \in [0,2\pi) \)} sampling a set of regularly spaced points in $D_k$, \textit{ii}) identifying points \( z_i \) \removetext{along each ray} where $d_k(z_i) = d_W$, and \textit{iii}) mapping them to $W_k$ via  \( \phi_k \). Repeating this process for multiple distance levels partitions $W_k$ into nested, conformal ribbons (Fig. \ref{fig:mappingAlgorithm}--\textit{e})}.

\removetext{Let $W$ represent a footprint partition, and $d : W \rightarrow \mathbb{R}$ denote the distance function, which measures the distance from a point\(w \in W \) to the nearest boundary point. Let $\phi_k : D_k \rightarrow W_k$ be a Schwarz-Christoffel conformal map that projects the unit disk $D_k$ onto the footprint subregion $W_k$. To simplify computations,
we define the pull-back of the distance function to the disk, $d_k : D_k \rightarrow \mathbb{R}$, as:}

\removetext{
\vspace{-3mm}
\begin{equation}
    d_k(z) = d(\phi_k(z)), \quad z \in D_k.
\end{equation}
\vspace{-5mm}
}

\removetext{\noindent Thus, the pull-back $d_k$ assigns to each point \( z \in D_k \) the distance from its mapped image \( \phi_k(z) \) in the footprint to the nearest boundary point in $W$. 
This formulation enables the computation of level sets of the distance function (\textit{i.e.}, contours where $d_k(z)$ remains constant) within the regular parametric disk domain rather than the irregular footprint geometry. 
To construct a discrete contour, we sample a set of radial directions \( \theta_i \in [0,2\pi) \) over $D_k$. Along each ray, we query for the points \( z_i \) satisfying $d_k(z_i) = d_W$, where $d_W$ denotes a target distance value. The coordinates $z_i$ are subsequently mapped through \( \phi_k \) to their corresponding positions $ w_i \in W_k$.
Interpolation over the mapped samples yields the desired contour in the original footprint domain.
By repeating this process across multiple prescribed distance levels, the resulting mapped contours partition $W_k$ into nested, conformal ribbons.
}

\vspace{1mm}
\noindent \highlight{\textbf{S$_5$.} \textit{Sector construction.} To define a sector in the footprint subregion between two consecutive boundary points $w_1, w_2 \in W_k$, we compute their pre-images $z_i = \phi_k^{-1}(w_i) $ for $i=1,2$ in the parametric disk $D_k$.
Expressing each $z_i$ in polar coordinates yields angular values  $\theta_1$ and $\theta_2$, which define two radial directions from the disk center.
Sampling points along these directions and mapping them back to $W_k$ via $\phi_k$ produces the sector curves ((Fig. \ref{fig:mappingAlgorithm}--\textit{f}).}

\removetext{To define a sector within the footprint domain $W_k$ based on two consecutive boundary points $w_1, w_2 \in W_k$, we compute their pre-images in the parametric domain $D_k$.
Since Schwarz-Christoffel maps are smooth bijections, we use the inverse conformal map to obtain  $z_i = \phi_k^{-1}(w_i) $ for $i=1,2$. 
Each pre-image $z_i$ is then expressed in polar coordinates, yielding an angular coordinate
$\theta_i$ relative to the center of the disk. The two angles $\theta_1$ and $\theta_2$ define radial directions emanating from the center, along which we sample discrete points. 
These points are subsequently mapped forward through $\phi_k$ to generate two curves in $W_k$, corresponding to the boundaries of the desired sector. This construction ensures that
the resulting sectors preserve the conformal structure of the mapping and maintain consistent angular relationships within the footprint, thereby minimizing the introduction of graphical artifacts in the embedded lens visualization.}

\vspace{1mm}
\noindent \textcolor{black}{\textbf{S$_6$.} \textit{Level set stitching.}} Finally, we \removetext{must seamlessly} merge the conformal maps defined over individual graph nodes into a unified geometric layout. \removetext{To this end, we join the} This is achieved by joining level set segments across adjacent nodes \removetext{by identifying} via matching points along shared boundaries.
Since adjacent nodes share common edges in the footprint, \removetext{there always
exist points where} their corresponding level sets intersect at common points, ensuring a continuous and coherent conformal layout across the entire footprint.\removetext{, as depicted in Figure \ref{fig:mappingResults}.}

\section{Visualization}
\label{sec:visualization}
Our lens visualization aims to facilitate spatiotemporal analysis across building facades; thus, after computing a conformal layout via our mapping algorithm, temporal displays are overlaid onto the building. In this section, we articulate the domain and design objectives guiding our approach, and present the rationale underlying the visual and interaction designs aligned with these objectives.

\subsection{Domain and Design Goals}
\label{subsec:designGoals}
Prior research has identified the following spatiotemporal analysis tasks, {T$_i$}, as commonly performed in 3D urban spaces \cite{mota2022comparison, miranda2024state}:

\vspace{1mm}
\noindent \textbf{T$_1$.} \textit{Spatial comparison across building facades}. 
In 3D city models, visual analyses are often driven by spatial rather than temporal dimensions. In this sense, \textit{spatial comparison} tasks involve multiple known spatial locations while investigating unknown temporal values. For instance, \removetext{such} tasks may involve questions like: ``how do shadow behaviors differ across various facades'' or ``which building facade receives the most shadow during winter?''

\vspace{1mm}
\noindent \textbf{T$_2$.} \textit{Temporal comparison in a building facade}. Another common task is \textit{temporal comparison}, where multiple time instances are assessed at a single known spatial location. For instance, one might compare shadow behavior between summer and winter on a given facade: ``in which season does the facade receive the least shadow?''

\vspace{1mm}
\noindent Building on the aforementioned analysis tasks, we established the following design goals, {DG$_i$}, for our immersive lens visualization:

\vspace{1mm}
\noindent \textbf{DG$_1$.} \textit{Anchor temporal information to spatial referents}. Since the tasks require associating temporal values with their corresponding spatial referents, our conformal mapping algorithm is used to carefully overlay temporal displays onto the building, thereby enabling in-place analysis with seamless coherence across space and time.

\vspace{1mm}
\noindent \textbf{DG$_2$.} \textit{Mitigate spatial displacement}. While embedding temporal displays, it is desirable to minimize disruption to the adjacent space in order to preserve users' sense of spatial orientation, which has been identified as an essential factor in 3D urban data analysis \cite{chen2021urbanrama}. To this end, we design an x-ray de-occlusion method that not only provides a clear line of sight to the temporal plot but also minimizes alterations to the building structure and its surroundings.

\vspace{1mm}
\noindent \textbf{DG$_3$.} \textit{Favor the visualization of temporal trends and extremes}. 
Domain experts often examine \textit{extreme} temporal values (\textit{e.g.}, ``which facade receives the most shadows during summer?''), as well as temporal \textit{trends} (\textit{e.g.}, ``is there minimal variation in solar incidence on a given facade throughout the day?'') \cite{mota2022comparison}. 
Accordingly, our visual design \removetext{places} favors enabling the perception of overall trends and extremes, rather than representing specific numeral values.

\vspace{1mm}
\noindent \textbf{DG$_4$.} \textit{Design for simplicity and intuitiveness}. 
Since we do not assume background knowledge or skill levels among users, our designs are guided by intuitiveness and simplicity. For this purpose, we considered a range of design choices that leverage human cognitive and motor abilities, as well as real-world knowledge. For instance, our de-occlusion design is a view-dependent cutaway inspired by well-known architectural illustrations, which makes use of humans’ natural cognitive ability to mentally reconstruct missing spatial structures. 
\removetext{Furthermore, this design takes advantage of motor coordination by enabling multiple simultaneous actions: continuous head and full-body motions control the cutaway perspective, while the hands remain available for complementary interactions.}


\begin{figure}[t!]
    \centering
    \includegraphics[width=1\linewidth,keepaspectratio=true]{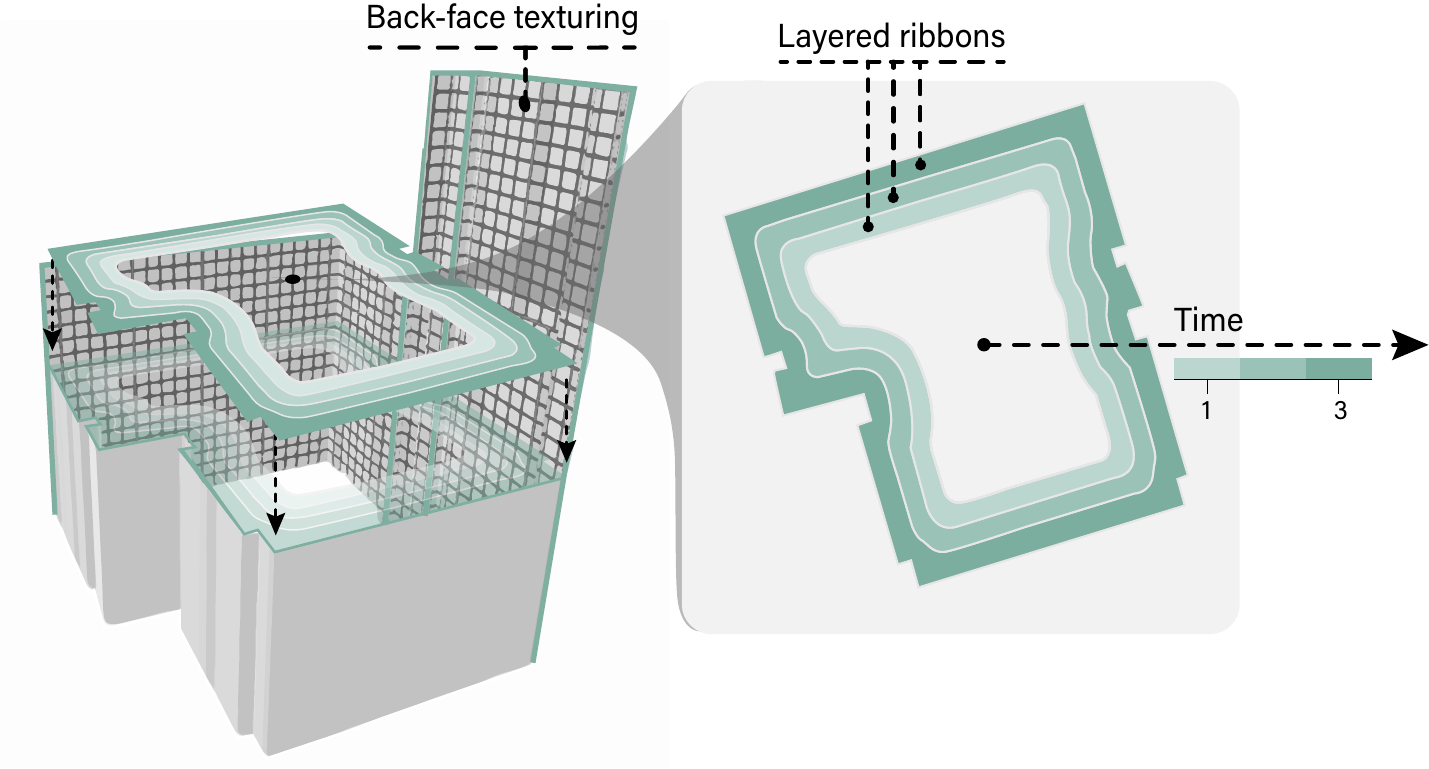}
    \caption{Double-sided cutaway rendering (\textit{left}) and conformal lens layout with layered ribbons encoding distinct time instances (\textit{right}).}
    \label{fig:timeDirection}
\end{figure}

\begin{figure}[t!]
    \centering
    \includegraphics[width=1\linewidth,keepaspectratio=true]{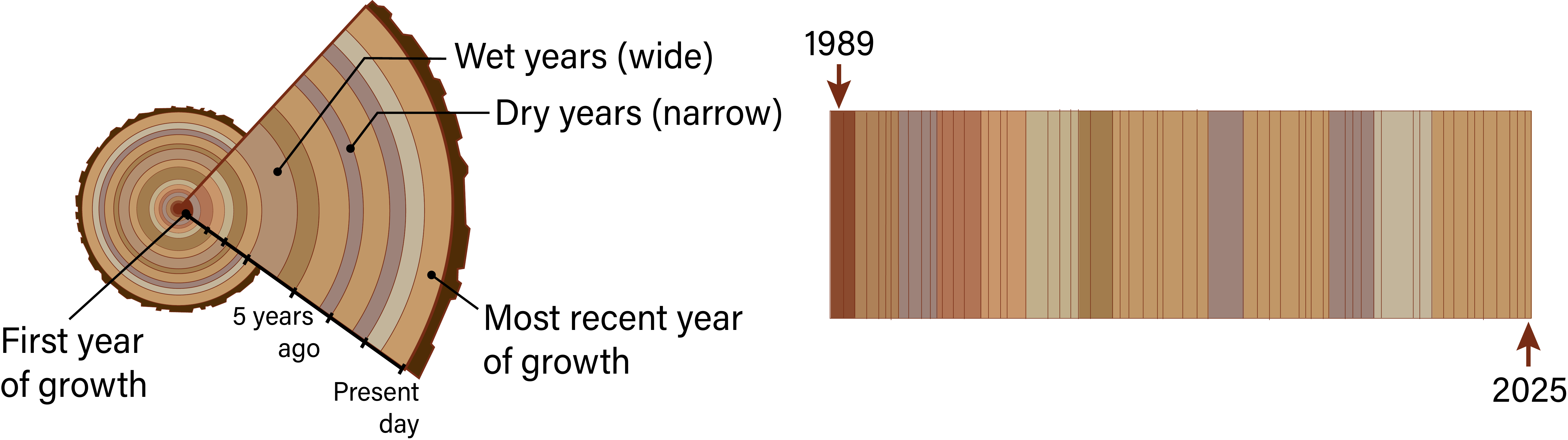}
    \caption{Tree-ring chronology (\textit{left}) and a cross-sectional view illustrating high-density temporal data (\textit{right}). 
    }
    \label{fig:treeRing}
\end{figure}

\subsection{Visualization Design}
Guided by our design goals, we addressed the subsequent design considerations in our spatiotemporal lens visualization.

\begin{figure}[b!]
    \centering
    \vspace{-5mm}
    \includegraphics[width=1\linewidth,keepaspectratio=true]{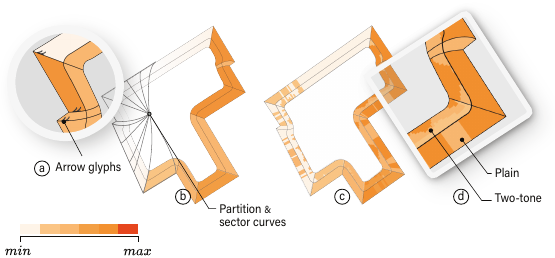}
    \caption{Color-coded lens visualizations using coarse (b) and fine (c) resolutions. Arrow glyphs aligned with section curves convey temporal progression (a). Two alternatives for encoding time-varying attribute values along ribbons are shown: plain color-coding is more space-efficient, and two-tone pseudocoloring enhances perceptual precision (d).}
    \label{fig:encodings}
\end{figure}

\begin{figure}[b!]
    \centering
    \includegraphics[width=1\linewidth]{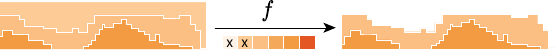}
    \caption{\highlight{Illustration of a lens function that modulates the visual prominence of selected attribute ranges in a two-tone pseudo coloring: low-value ranges are suppressed to visually accentuate high-value regions.}}
    \label{fig:encodingsFilter}
\end{figure}

\vspace{1mm}
\noindent \textbf{Spatial de-occlusion.} A major design decision concerns how to address occlusion in the building to ensure a clear line of sight to the embedded temporal display. Drawing inspiration from traditional illustrative architectural visualizations, we employ \textit{cutaways}, as they preserve the overall outer layout of the building (\textbf{DG$_2$}) while leveraging human cognitive capability to mentally reconstruct the spatial structures that have been cut away (\textbf{DG$_4$}) \cite{konev2018fast}.
Cutaways fall within the family of x-ray de-occlusion techniques, which inherently offer limited spatial depth cues, ultimately hindering the perception of the spatial correlation between the temporal display and the building's surface layers. Although internal shadows could provide strong disambiguation cues, we chose to omit them to prevent potential interference with the perception of color-coded temporal values. Instead, we apply distinct visual styling to the inner and outer building surfaces, which not only accentuates the prominence of the cut but also provides sufficient spatial cues for users to comprehend the spatial arrangement, as illustrated in Fig.~\ref{fig:timeDirection}--\textit{left}.
Furthermore, our cutaway is designed to be \textit{view-dependent} (\textbf{DG$_4$}). In conventional focus-and-context/lens visualizations, viewpoint dependency is often a marginal consideration, as traditional displays typically occupy users’ full attention, with shifts in focus occurring only when users explicitly select a new area of interest.
In contrast, within immersive environments, user-selected foci may remain relevant even as users physically navigate around the space. Hence, a view-dependent de-occlusion design ensures continuous visibility of areas of interest by enabling lenses to better anticipate and align with user intent, thus reducing the need for manual input.
This dynamic response leverages innate human motor capabilities to regulate concurrent actions: head and full-body movements regulate the lens visibility, while the hands remain free for complementary lens interactions (\textbf{DG$_4$}).

\vspace{1mm}
\noindent \textbf{Temporal direction.} Embedding temporal displays onto buildings introduces the challenge of conveying the direction of time progression. While conventional orientations used in linear axis-based visualizations (\textit{e.g.}, where time advances from left to right or bottom to top according to a Cartesian coordinate system) are intuitive, they do not directly apply to our specific case of conformal, spatially-embedded visual layouts.
Radial layouts are also commonly employed in abstract visualizations, offering three main strategies for conveying time direction: ring-based, space-filling, and polar-based encodings \cite{filipov2021gone}. In the polar-based design, the center point represents the origin of the coordinate system, and the distance between any point and the center encodes a temporal semantic meaning. Time progression is thereby represented as rays directed outward from or near the center. Polar-based layouts are intuitive, as they mirror many temporal phenomena observed in nature, \textit{e.g.}, ripples in water, shockwaves, and tree rings, as shown in Fig.~\ref{fig:treeRing} (\textbf{DG$_4$}).
Hence, we adopt a \textit{polar-based time axis} not only for its widespread use and intuitive comprehensibility but also because it arranges temporal attribute values linearly and in close proximity to their spatial referents (\textbf{DG$_1$}).
Specifically, our temporal display is designed around the concept of \textit{nested, conformal ribbons}, each representing a distinct time instance, with temporal direction encoded through the appropriate polar-based ordering of the ribbons, as illustrated in Fig.~\ref{fig:timeDirection}--\textit{right}.

While we believe that our real-world analogy effectively conveys chronological order, one might consider overlaying additional visual cues to further reinforce the temporal reference frame and users' mental models. In this sense, visual elements such as labels and symbols are commonly used to disambiguate the direction of time; among these, \textit{arrow glyphs} are widely employed in both abstract and spatial visualizations.
Although it is well established that glyph sampling strategies significantly affect both clutter and the perception of encoded information \cite{borgo2013glyph}, an in-depth discussion of this topic is beyond the scope of this paper. Nevertheless, we posit that minimal, directional glyphs can be strategically placed and oriented according to the curvature of footprint sector boundaries, using the curves computed during our mapping algorithm. This alignment may enhance the legibility of the underlying footprint layout while minimizing interference with visual variables dedicated to temporal encoding (see Fig.~\ref{fig:encodings}--\textit{a}).

\begin{figure*}[t!]
    \centering
    \includegraphics[width=1\linewidth]{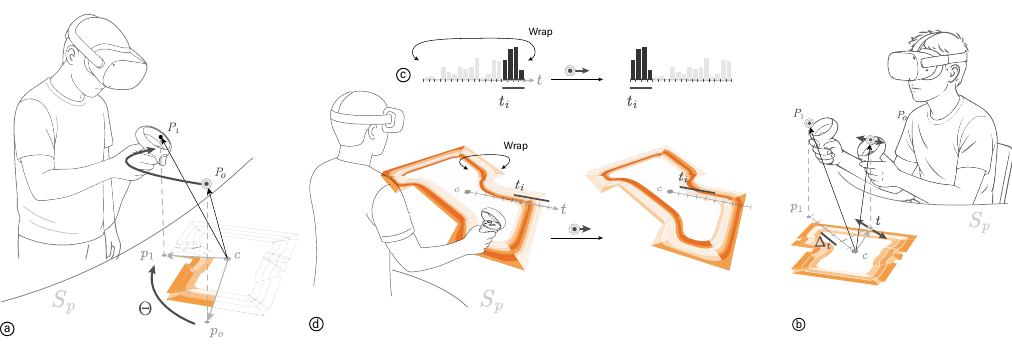}
    \caption{Illustrations of lens interactions. (a) An angular drag gesture \( \overrightarrow{P_0P_1} \) (\iconDrag) with either hand projects the drag trajectory onto the virtual table $S_p$, performing spatial filtering by bringing the corresponding footprint sectors into focus. (b) A trigger press at position $P_1$ (\iconPress) reveals the temporal axis and sets its endpoint $p_1$ on the surface of the table. A subsequent radial drag initiated at $P_0$ (\iconDrag) with either hand enables temporal navigation along the time axis, displaying distinct time intervals \( \Delta_t \) within the lens image. (c, d) Exemplary comparison of \textit{wrapped} temporal reordering using drag gestures (\iconDrag) with cyclic data, shown on a bar chart and in our lens visualization. This wrapping interaction aims to enhance the readability and discovery of temporal trends, such as a monotonically increasing sequence of temporal values along adjacent ribbons.}
    \label{fig:interactions}
\end{figure*}

\vspace{1mm}
\noindent \textbf{Temporal encoding.} Another design consideration involves determining appropriate visual encodings for representing temporal attributes. Given that domain experts often examine \textit{scalar} temporal functions defined on the building geometry, our design ideations favor representing this data type.
One straightforward option is to employ color mapping, wherein temporal attribute values are mapped to the color saturation channel, an encoding widely adopted in 3D urban data visualizations, as shown in Fig.~\ref{fig:encodings}--\textit{b, d} \cite{mota2022comparison}. 
While it is well known that humans exhibit limited perception and accuracy when extracting quantitative information from colors, color remains a familiar and adequate representation for conveying general trends and extrema (\textbf{DG$_3$}).
Nonetheless, to enhance graphical perception, we can consider other space-efficient color-coded approaches. For instance, techniques such as two-tone pseudo coloring \cite{saito2005two, tominski2012stacking} enable more precise value discrimination by compressing information spatially, as depicted in Fig.~\ref{fig:encodings}--\textit{c, d}. However, this method is sensitive to the thickness of the visual substrate, \textit{i.e.}, when ribbon thickness is reduced excessively, perceptual accuracy may deteriorate significantly.
To improve color discrimination, one could enable on-demand lens adjustments that attenuate the visual prominence of selected attribute ranges, thereby allowing users to visually emphasize or suppress temporal features of interest (Fig.~\ref{fig:encodingsFilter}).




 \begin{figure*}[t!]
     \centering
     \includegraphics[width=1\linewidth]{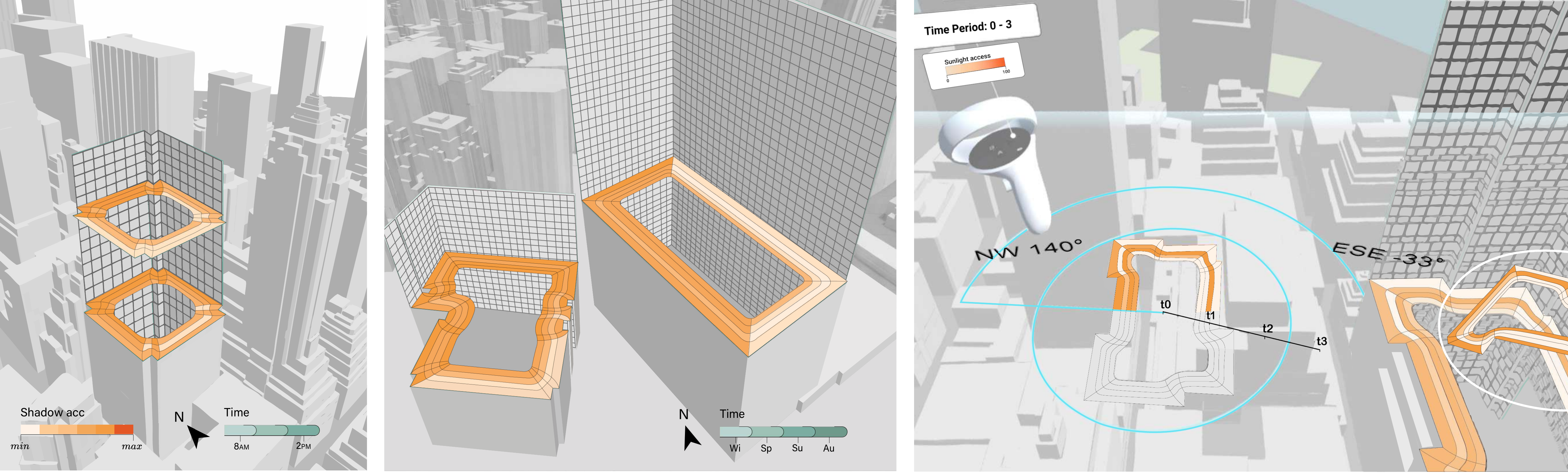}
     \caption{\highlight{\textit{Left} and \textit{middle}: Application scenarios demonstrating lens use to inspect shadow accumulation across daily time windows and yearly seasons.} \textit{Right}: Lens replicas distributed over the semi-transparent virtual table. The cyan circle indicates the active lens interaction space, within which the user executes an angular drag to query spatial sectors spanning directions from ESE $-33^\circ$ to NW $140^\circ$.
     }
     \label{fig:applicationScenarios}
 \end{figure*}
 
\vspace{-1mm}
\subsection{Interaction Design}
\vspace{-1mm}
In recent years, urban data has emerged as a promising application domain for immersive analytics due to its inherent spatial complexity and the demand for collaborative analysis \cite{Chen2017}.
Accordingly, urban visualizations have begun to leverage the unique affordances of immersive technologies: \textit{i}) wider stereoscopic visual space, \textit{ii}) varied spatial interaction modalities, \textit{e.g.} mid-air gestures and tangibles, \textit{iii}) allocentric and egocentric spatial perspectives, \textit{iv}) and co-located or distributed collaborative environments \cite{Zhang2021urbanvr, chen2021urbanrama, ens2020uplift, lock2019holocity, alonso2018cityscope, elvezio2018collaborative, wu2017efficient, chen2017immersive}.
Therefore, although our lens remains operable on conventional displays, it was deliberately designed to support spatiotemporal urban data analysis within immersive environments, particularly leveraging egocentric perspectives and mid-air spatial interactions.

\vspace{1mm}
\noindent \textbf{Global interaction space.} In our immersive interaction design, we distinguish between \textit{global} and \textit{local} interaction spaces. For the former, we incorporate a \textit{virtual desk} metaphor in which a virtual reproduction of a work desk remains within arm's reach in front of the user \cite{wagner2018virtualdesk}.  \highlight{This design provides a versatile workspace where users can either walk around or sit down while interacting with content positioned on the table surface (Figs.~\ref{fig:interactions} and \ref{fig:applicationScenarios}--\textit{right}).
This metaphor \removetext{reduces unnecessary large arm motions and head rotations in the wide immersive space, and} also enables direct manipulation of table contents via mid-air gestures that mirror familiar physical actions: users can reposition content via a trigger-press-and-drag gesture with either hand, and can scale content using a trigger-press-and-radial-drag gesture with both hands, dragging outward to enlarge and inward to reduce.}
These gestures leverage human real-world knowledge (\textbf{DG$_4$}), resulting in increased productivity and decreased task complexity through: \textit{i}) lower cognitive load for interaction learning and execution, thereby freeing mental resources for data interpretation, \textit{ii}) bimanual interactions, as users' familiarity with physical actions facilitates the combination of multiple operations, and \textit{iii}) reduced context switching, allowing eyes-free interaction while maintaining attention on the task at hand \cite{jacob2008reality, hutchins1986direct}.

\highlight{The virtual table contains user controls and compact \textit{2D replicas} of the corresponding 3D lens images.
The replicas resemble a world-in-miniature, a metaphor widely used in immersive environments due to two core affordances \textcolor{red}{\cite{stoakley1995virtual, danyluk2021design}}.
First, the replicas provide a scaled-down exocentric view that complements the large-scale egocentric virtual environment.
By default, the placement of replicas on the table preserves their relative spatial positioning to facilitate spatial orientation and cognitive mapping between the 2D representations and their 3D counterparts.}
\highlight{Second, the replicas serve as \textit{local} interaction spaces at a reduced scale within arm's reach, thereby mitigating large arm motions and head rotations that can reduce mid-air interaction precision and increase fatigue in wide immersive environments \textcolor{red}{\cite{mendes2019survey, iqbal2021reducing}}}.

\vspace{1mm}
\noindent \textbf{Local lens interaction space.} Tapping a lens replica with either hand activates its local, \textit{two-dimensional} interaction space. 
\highlight{The interaction design derives from the visual design of the layered temporal display: color variations along the radial axis encode temporal trends, while variations along the angular axis convey spatial patterns.
Similarly, \textit{radial} gestures, moving inward or outward from the lens center, interact with the \textit{temporal} dimension; \textit{angular} gestures around the center act on the \textit{spatial} dimension; and \textit{diagonal} gestures operate on both spatial and temporal dimensions.
The following lens interactions, marked as \textbf{I}$_i$, supplement the temporal plot to support the spatiotemporal tasks specified in Sec.~\ref{subsec:designGoals}.}

\vspace{1mm}
\noindent \textbf{I$_1$.} \textit{\textcolor{black}{Spatial filtering.}} To facilitate the spatial comparison of attribute behaviors (\textbf{T$_1$}), users can construct spatial filters using \textit{angular drag} gestures, by pressing the trigger button with either hand and dragging along the angular dimension, users direct visual focus to specific footprint sectors.
This gestural design not only resembles real-world actions, \textit{e.g.}, turning the head or upper body in angular sweeps to inspect a room, but also affords varying levels of precision depending on the distance from the center at which users perform the rotational gesture: gestures closer to the center cause coarse-grained selection, while those farther away enable finer-grained control over spatial filtering (\textbf{DG$_4$})\textcolor{black}{(see Fig.~\ref{fig:interactions}--\textit{a}}).

\vspace{2px}
\noindent \textbf{I$_2$.} \textit{\textcolor{black}{Temporal navigation.}}  For the analysis of temporal behavior (\textbf{T$_2$}), users can navigate along the temporal dimension by pressing the grip button with either hand and dragging along the radial axis, resembling universally recognizable gestures such as opening or closing arms to estimate size (\textbf{DG$_4$}). This time axis appears on demand, originating at the center of the lens and ending at a user-specified point to adjust its length (see \textcolor{black}{Fig.~\ref{fig:interactions}--\textit{b}}). Modifying the axis length directly influences the level of precision afforded by the \textit{radial drag} gesture.
Notably, users can perform both spatial and temporal filtering through a continuous drag gesture within the two-dimensional interaction space.

\vspace{2px}
\noindent \textbf{I$_3$.} \textit{\textcolor{black}{Temporal ordering.}} \removetext{Furthermore, }To support temporal comparison of cyclic data (\textbf{T$_2$}), we implement a wrapping ordering mechanism along the temporal dimension \textcolor{black}{\cite{chen2021rotate}}.
Here, \textit{wrapping} denotes the perceptual radial continuity achieved by connecting the inner and outer ribbons of our nested temporal display.
Compared to a conventional unwrapped axis, the wrapped representation offers two major advantages: \textit{i}) it centers the time interval of interest within the user's field of view, and \textit{ii}) it enables the continuous presentation of cyclic temporal data, eliminating visual discontinuities that would otherwise require users to mentally reassemble disjointed time segments to discern cyclical trends (see Fig.~\ref{fig:interactions}--\textit{c}).
Thus, users can carry out continuous temporal reordering via a \textit{radial drag gesture}, a drag outward from, or toward, the center of the lens image, as depicted in Fig.~\ref{fig:interactions}--\textit{d}. 

\vspace{2px}
\noindent \textbf{I$_4$.} \textit{\textcolor{black}{Attributive filtering.}}  To place visual focus on specific color-coded value ranges (useful for both spatial and temporal comparisons (\textbf{T$_1$}, \textbf{T$_2$})), our lens affords a function that modulates the visual prominence of specific value intervals, as illustrated in Fig.~\ref{fig:encodingsFilter}. This color-based filtering is triggered by tapping on the corresponding segments of the color legend in the virtual table.

\vspace{2px}
\noindent \textbf{I$_5$.} \textit{\textcolor{black}{Attributive ordering.}} Lens ribbons are chronologically ordered by default to convey temporal progression, but alternative criteria can also be useful, \textit{e.g.}, ordering ribbons by average temperature facilitates spatial comparison across facades (\textbf{T$_1$}) by revealing those exhibiting extreme thermal values or significant temperature disparities. To enable such attribute-based reordering, users can tap on the dropdown and toggle controls on the virtual table.


\vspace{-1mm}
\subsection{Application Scenarios}

\highlight{This section presents two application scenarios of our lens visualization for shadow analysis in New York City, shown in Fig.~\ref{fig:applicationScenarios} and linked to the spatiotemporal tasks described in Sec.~\ref{subsec:designGoals}.}

\vspace{1mm}
\noindent \highlight{\textit{Cross-height spatial comparison in building facades.}} \highlight{In dense urban environments, shade distribution can vary substantially along a building's height due to shadows cast by surrounding skyscrapers~\cite{moreira2023urban}.
To understand cross-height shadow behavior (\textbf{T$_1$}), we examine accumulated shadow over three separate time windows within a day (morning, afternoon, and mid-afternoon) on an office tower that rises 227 $m$ over 47 stories. The building is located in the Financial District, one of the densest areas of Lower Manhattan, and is surrounded by mid- to high-rise office buildings of comparable height.}
\highlight{By translating a lens across the building stories or by placing two or more lenses (Fig.~\ref{fig:applicationScenarios}--\textit{left}), we can compare the daily shadow behavior across facades at different heights:
In north-facing facades, shadow values remain consistently high throughout the day and across building heights, indicating low sensitivity to height.
In contrast, west- and south-facing facades exhibit the lowest shadow values from late morning to mid-afternoon. These facades show more pronounced height-dependent differences: higher stories experience lower shadow values than mid stories, indicating that even sun-favored facades at intermediate heights may remain shaded likely due to sun blocking from surrounding buildings.}

\vspace{1mm}
\noindent \highlight{\textit{Cross-season temporal comparison in building facades.}}
\highlight{Comparing shadow across seasons is relevant because seasonal changes substantially influence the amount of sunlight reaching a building facade~\cite{8283638}.
From a human health perspective, shadows offer thermal comfort during summer months, but prolonged seasonal shading reduce access to indoor sunlight and have adverse effects on occupant's mental health.
From a building energy perspective, seasonal shadow behavior directly affects heating and cooling demands, \textit{e.g.}, in summer, increased shading lowers indoor temperature and air-conditioning use.} 
\highlight{To understand cross-season shadow behavior (\textbf{T$_2$}), we examine accumulated shadows across the four seasons in two neighboring buildings near the Lower Manhattan waterfront: a large office tower with a height around 210 $m$ over 53 stories, and a shorter office building measuring around 110 $m$ over 27 stories.}
\highlight{By placing two lenses at comparable heights above ground on the buildings (Fig.~\ref{fig:applicationScenarios}--\textit{right}), we can observe that north facades have small seasonal variation and are severely shadowed throughout the year.}
\highlight{In contrast, south facades exhibit a strong seasonal swing with accumulated shadow values shifting significantly from higher levels in winter to lower levels in summer.
East-facing facades benefit from increased sunlight access provided by the open waterfront adjacency, exhibiting lower accumulated shadow than other facades.}
\highlight{However, the figure shows that during summer the shorter office building retains higher accumulated shadow, which may be attributed to shadowing from the taller tower situated slightly to the northeast.}
\section{Evaluation}
\label{sec:evaluation}
\highlight{This section presents a user study aimed at evaluating the effectiveness of our immersive lens visualization.} 
It followed a between-subjects design, and employed an analysis scenario in which users were asked to perform the visualization tasks described in Sec.~\ref{subsec:designGoals}.
\highlight{This study was approved by the University of Calgary Research Ethics Board [REB21-1659\_REN3].}

\vspace{1mm}
\noindent \removetext{Selection and design of }{\textbf{\highlight{Visualizations.}} 
The literature describes a range of spatiotemporal and de-occlusion visualization designs applicable to urban data analysis. To ensure a tractable scope of our study, we narrow this set based on the following criteria: the selected techniques should \textit{i}) be in widespread use within the urban domain, \textit{ii}) rely on the same visual encoding channels, and \textit{iii}) provide an equivalent level of interactivity. Accordingly, we compare our visualization with commonly used \textit{interchangeable} and \textit{x-ray} techniques, whose combinations yield the following design conditions, each labeled as \textbf{V}\textit{\small is}\textbf{$_i$} \highlight{and illustrated in Fig.~\ref{fig:studyVis}}:
\noindent \textbf{V}\textit{\small is}\textbf{$_1$.} \visLabel{TemporalJX} (baseline) displays one data instance at a time, and users navigate between time steps using the right thumbstick. A sequential colormap encodes the time-dependent quantitative attribute.
\noindent \textbf{V}\textit{\small is}\textbf{$_2$.} \visLabel{TemporalJX+Xray} reveals data across all building facades but does not support the simultaneous display of different time instances. It employs the same color encoding and time-switching interaction as in the baseline condition. 
\noindent \textbf{V}\textit{\small is}\textbf{$_3$.} \visLabel{EmbeddedV+Xray} displays data across all sides of the building, presenting all data instances simultaneously. It uses the same color encoding as the other conditions.

\vspace{1mm}
\noindent \highlight{\textbf{Interactions.} The visualization designs supported equivalent interactivity: \textit{i}) camera control via head and body motions; \textit{ii}) switching time via the right thumbstick, only \textbf{V}\textit{\small is}\textbf{$_1$} and \textbf{V}\textit{\small is}\textbf{$_2$}; and \textit{iii}) selecting and submitting the task answer in the UI via the left thumbstick and trigger, respectively. Lens replicas and associated interactions were not used. To isolate the effects of the visualization designs, we limited interactions to a minimal set, thereby avoiding confounding effects from replica-based interactivity.}

\vspace{1mm}
\noindent  \textcolor{black}{\textbf{Datasets.}} We used the public Manhattan OSM model \cite{manhattanModel}, with a temporal attribute defined on building facades representing accumulated sunlight access over eight hours in each of the four yearly seasons. To increase data variability, we added random noise.

\vspace{1mm}
\noindent \textbf{Participants.} We recruited a total of 28 participants (\textcolor{black}{8 female and 20 male}) \highlight{across the visualization conditions: \textbf{V}\textit{\small is}\textbf{$_1$}: 9, \textbf{V}\textit{\small is}\textbf{$_2$}: 9, and \textbf{V}\textit{\small is}\textbf{$_3$}: 10.} None reported any color vision deficiency, and all had normal or corrected-to-normal vision. Ages ranged from 18 to 36 (avg. = 27.18, std. dev. = 5.6). \highlight{Most participants were students (23/28) in Computer Science or Computer Engineering programs, with prior use of 3D visualization (22/28) and VR (17/28)}. All participants \highlight{provided informed consent prior to participation,} were volunteers and received no monetary compensation.

\vspace{1mm}
\noindent \textbf{Apparatus.} The visualizations were implemented using Unity 3D and an Oculus Quest 2 (1832x1920 display resolution per eye). We obtain interactive frame rates of 45 FPS using a laptop with an Intel®Core ™i7 processor and GeForce RTX 3060 GPU.

\vspace{1mm}
\noindent \textbf{Procedure.} Each session consisted of three parts: an introduction and training phase, the main trials, and a post-study questionnaire, lasting a total of 60–90 minutes. During the introduction and training phase, we explained the purpose of the experiment and the visualization to be used. Participants then answered three training trials. In the main trials, they answered six questions \removetext{and} without receiving feedback on the correctness of their responses. They were asked to respond as quickly as possible without compromising accuracy.

\begin{figure}[t!]
    \centering
    \includegraphics[width=1\linewidth,keepaspectratio=true]{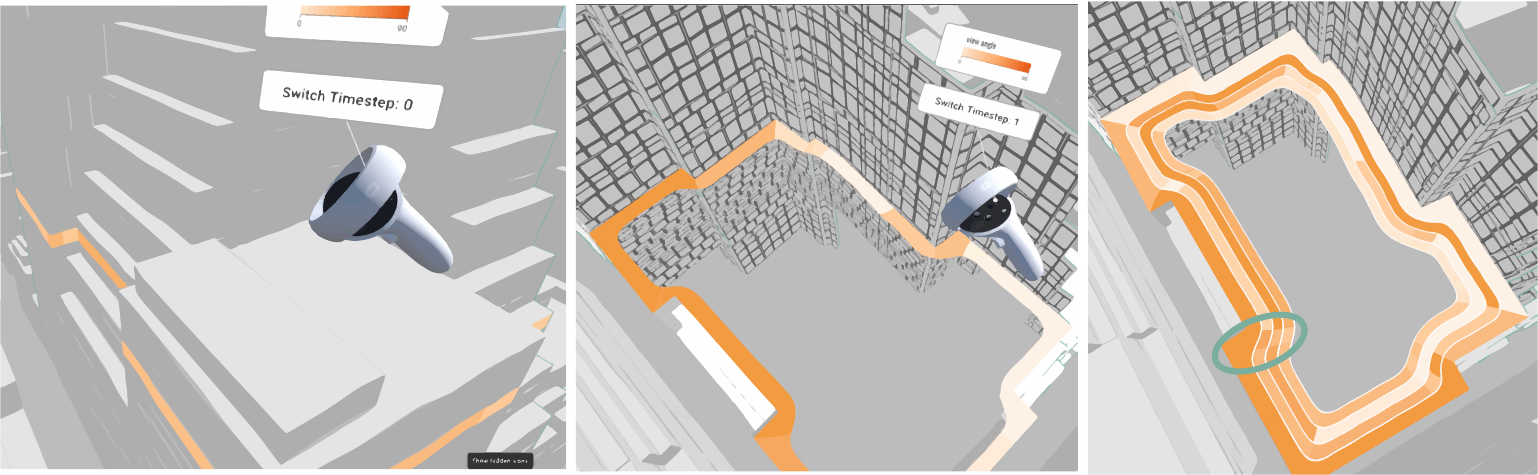}
    \caption{\highlight{Visualization designs in a sample study task. From \textit{left} to \textit{right}: \visLabel{TemporalJX}, \visLabel{TemporalJX+Xray}, and \visLabel{EmbeddedV+Xray}. The \textcolor{meanColor}{green circle} (right) highlights the facade with the longest continuous attribute increase, corresponding to the task answer $[t_0$- $t_3]$.}
    }
    \label{fig:studyVis}
\end{figure}

\vspace{1mm}
\noindent \textbf{Task.} The purpose of our study was to evaluate the effectiveness of our visualization \highlight{design} in supporting the analysis of \textit{self-occluded} and \textit{time-varying} data.
\highlight{Accordingly, we addressed these aspects as part of a single, integrated task that required participants to \textit{i}) access partially or fully occluded data, and \textit{ii}) relate data variations over time. Participants were asked\footnote{This task falls under the \textit{temporal comparison of trends at the building story level} category in a recent 3D urban visualization task taxonomy \cite{mota2022comparison}.}: ``\textit{In the reference building, during which time range does the attribute have its longest continuous growth?}".
\highlight{Figure \ref{fig:studyVis}} shows a sample task: participants would first access each facade individually to identify, for that facade, the time range exhibiting the longest continuous increase in attribute values; they would then compare these facade-specific time ranges and select the longest overall range as their final answer.}

\vspace{1mm}
\noindent \textbf{Measures.} For each trial, we \highlight{collected data} for two performance metrics: \textit{i}) task \textit{completion time}, measured from the moment participants viewed the trial screen
until they submitted a response; and \textit{ii}) \textit{error}, defined as $e = 1 - $ IoU, where IoU refers to the intersection over union method and quantifies the overlap between the ground truth and the participant’s selected time interval. Thus, higher values indicate a greater temporal mismatch. Additionally, we collected throughout each trial the participant's \textit{confidence level} as a subjective metric, and tracked their \textit{head pose} as a physical measure. During each trial, we also logged qualitative observations related to participants’ behaviors, interaction strategies, and aloud comments. After completing all trials, participants filled out a post-study questionnaire to share both positive and negative opinions, as well as ideas for improving the visualization.

\vspace{1mm}
\noindent \highlight{\textbf{Hypotheses.} Given the four aforementioned measures, we formulated four hypotheses to guide our analysis
:
\noindent \textbf{H}\textbf{$_{pos}$.} \visLabel{EmbeddedV+Xray} and \visLabel{TemporalJX+Xray} may impose less \textit{head pose} changes than \visLabel{TemporalJX}, as the x-ray de-occlusion uncovers facade-wise data with reduced need for physical motion.
\noindent \textbf{H}\textbf{$_{ct}$.} \textit{Completion time} may exhibit a similar behavior, increasing according to the amount of required physical movement and viewpoint changes.
\noindent \textbf{H}\textbf{$_{err}$.} \textit{Error}  may vary with the degree to which data is simultaneously visible, with higher accuracy associated with \visLabel{EmbeddedV+Xray}, followed by \visLabel{TemporalJX+Xray}, and lowest with \visLabel{TemporalJX}.
\textbf{H}\textbf{$_{conf}$.} \textit{Confidence level}  may vary similarly, with confidence reflecting the degree of simultaneous visibility of data.}

\begin{figure*}[ht!]
    \centering
    \includegraphics[width=1\linewidth,keepaspectratio=true]{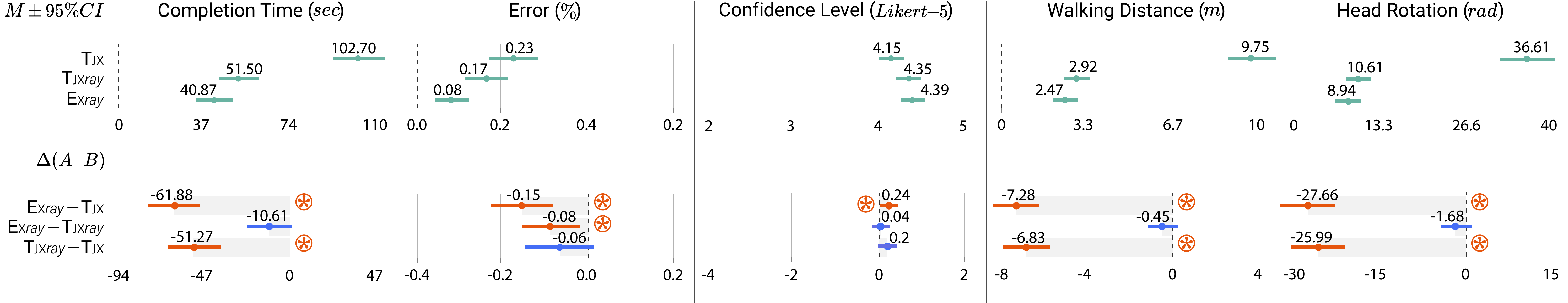}
    \caption{ Results for \textcolor{meanColor}{Mean} (\textit{top}) and \textcolor{pairwiseColor}{pairwise differences} (\textit{bottom}) across the three visualization conditions: \visLabel{EmbeddedV+Xray} ($\mathrm{E}_{\mathrm{X}\mathit{ray}}$), \visLabel{TemporalJX+Xray} ($\mathrm{T}_{\mathrm{JX}\mathit{ray}}$), and \visLabel{TemporalJX} ($\mathrm{T}_{\mathrm{JX}}$). From \textit{left} to \textit{right}, columns report completion time ($sec$), error ($\%$), confidence level (Likert-5 scale), walking distance ($m$), and head rotation ($rad$) across all tasks. Error bars represent 95\% Bootstrap CIs. Evidence of differences is marked in \textcolor{evidenceColor}{orange} along with an asterisk \iconAsterisk\ (the farther from zero and the narrower the CI, the stronger the evidence).}
    \label{fig:results}
\end{figure*}

\section{Results}
\label{sec:results}

We now report and interpret findings for task \textit{{completion time}}, \textit{{error}}, \textit{{confidence level}}, and \textit{{head pose}} from a total of \textcolor{black}{168} trials, as illustrated in Fig.~\ref{fig:results}. Our reporting methodology uses estimation techniques and reports sample means with confidence intervals (CI) rather than \textit{p}-value statistics \cite{cumming2005, dragicevic2016}, following recommended practices \cite{american2019}. However, a \textit{p}-value approach can always be obtained from the data by following similar techniques. For our inferential analysis, we use pairwise differences between means and their 95\% CIs\footnote{A CI of \textit{differences} that does not cross zero provides evidence of differences. The further away from zero and the smaller the CI, the stronger the evidence.\label{footnote}}, indicating the range of plausible values for the population mean. We use BCa bootstrapping to construct the confidence intervals, and refer to the supplementary material for the data files.

\noindent \textbf{Completion time} Participants spent on average less time using \visLabel{EmbeddedV+Xray} (40.87s, CI = [33.05, 49.25]), and slightly longer duration using \visLabel{TemporalJX+Xray} (51.5s, [43.24, 60.24]). Task completion time increased substantially with \visLabel{TemporalJX} (102.7s, [91.7, 114.4]). Pairwise comparisons provide strong evidence that \visLabel{TemporalJX} was significantly slower than the other visualizations, with an average difference of 56.57s (CI = [42.28, 70.93]). In contrast, no clear evidence of a time difference was found between the two x-ray-based visualizations, \textit{i.e.} \visLabel{EmbeddedV+Xray} and \visLabel{TemporalJX+Xray}.

\noindent \textbf{Error} Participants’ mean error was lowest with \visLabel{EmbeddedV+Xray} at 8.17\% ([4.47\%, 12.34\%]), increased with \visLabel{TemporalJX+Xray} at 16.66\% ([11.57\%, 22.06\%]), and reached 23.3\% ([17.43\%, 29.16\%]) with \visLabel{TemporalJX}. Pairwise comparisons reveal significant differences between our technique and the two temporal juxtaposition-based visualizations: \visLabel{EmbeddedV+Xray} was on average 11.83\% ([4.93\%, 18.59\%]) more accurate than both \visLabel{TemporalJX+Xray} and \visLabel{TemporalJX}.

\noindent \textbf{Confidence level} In the baseline condition \visLabel{TemporalJX}, the average \removetext{slightly} confidence level was slightly lower at 4.15 (CI = [4.0, 4.29]) compared to \visLabel{TemporalJX+Xray} (4.35, [4.2, 4.5]) and \visLabel{EmbeddedV+Xray} (4.39, [4.26, 4.54]). Pairwise comparisons indicate a small but significant difference in confidence with \visLabel{EmbeddedV+Xray} over \visLabel{TemporalJX}, with an average difference of 0.24 points ([0.04, 0.44]). No differences were found between the other visualization pairs.

\noindent \textbf{Walking distance} Participants traveled an average of 2.47m ([2.0, 2.98]) using \visLabel{EmbeddedV+Xray}, and slightly more with \visLabel{TemporalJX+Xray} (2.92m, [2.42, 3.43]). A pronounced increase in travel distance was observed with \visLabel{TemporalJX}, where participants covered 9.75m on average ([8.83, 10.7]). Pairwise comparisons clearly show that \visLabel{TemporalJX} resulted in significantly greater travel distances than both \visLabel{TemporalJX+Xray} and \visLabel{EmbeddedV+Xray}.

\noindent \textbf{Head rotation} Participants performed an average of 8.94 radians ([7.13, 11.13]) of head rotation using \visLabel{EmbeddedV+Xray}, and marginally more with \visLabel{TemporalJX+Xray} (10.61 rad, [8.81, 12.52]). A substantial growth was observed with \visLabel{TemporalJX}, where head rotation averaged 36.61 radians ([32.39, 40.88]). Pairwise comparisons reveal clear differences between x-ray-based visualizations and \visLabel{TemporalJX}, while no meaningful difference was found between \visLabel{EmbeddedV+Xray} and \visLabel{TemporalJX+Xray}.

\begin{figure}[t!]
    \centering
    \includegraphics[width=\linewidth,keepaspectratio=true]{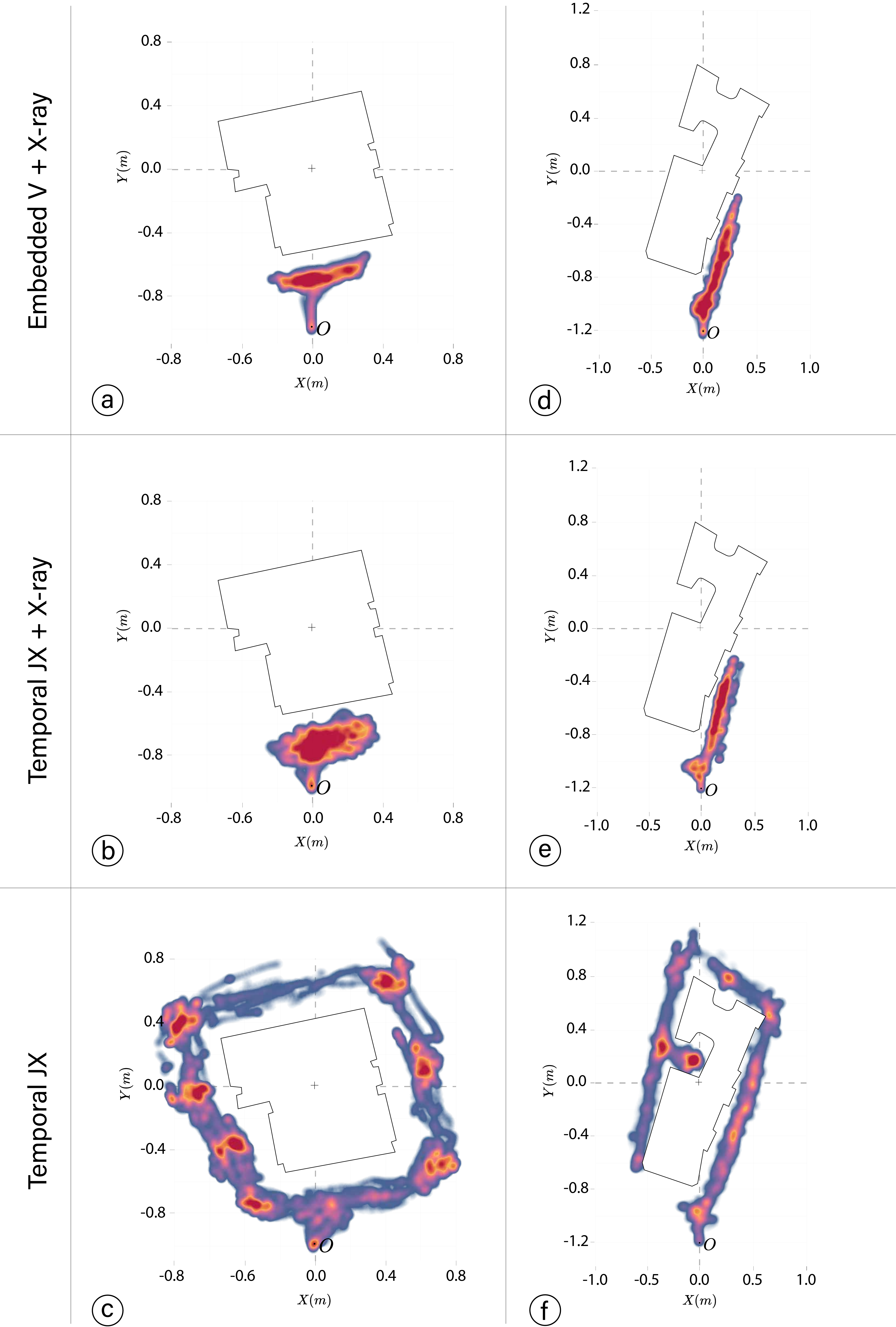}
    \caption{Density maps showing participants' positions across visualizations (\textit{rows}) for two tasks (\textit{columns}) using a sunset diverging colormap, which transitions from cool violet-blues to warm oranges and red tones. The origin $O$ represents the start position. 
    }
    \vspace{-5mm}
    \label{fig:resultsDensity}
\end{figure}
\section{Discussion}
\label{sec:discussion}

This section discusses not only the quantitative performance differences identified among the three visual designs, but also integrates our own observations, questionnaire responses, and tracked head pose data to provide a more comprehensive characterization of participants' behaviors and strategies when using each visualization. We catalog these findings as a set of design recommendations, {DR$_i$},  intended to inform both visualization researchers and practitioners. \removetext{across urban and broader application domains.}

\vspace{1mm}
\noindent \textbf{DR$_1$.} \textit{Integrate x-ray de-occlusion to reduce spatial navigation, thereby accelerating task completion.} We saw significant differences in completion time, walking distance, and head rotation between \visLabel{TemporalJX} and the other visualizations, but no clear difference between \visLabel{EmbeddedV+Xray} and \visLabel{TemporalJX+Xray}. This could indicate that the positive performance and physical results may be due to their x-ray capability.
\highlight{The density maps in Fig.~\ref{fig:resultsDensity} show similar navigation behaviors for \visLabel{EmbeddedV+Xray} and \visLabel{TemporalJX+Xray}, elucidating the benefits of x-ray de-occlusion: participants took only a few steps forward during the task, which seems to be sufficient to obtain a satisfactory overview of the data across all building facades. Consequently, participants exhibited significantly less physical movement and required considerably less time to respond when using x-ray-based visualizations compared to \visLabel{TemporalJX}}
\highlight{, confirming \textbf{H}\textbf{$_{pos}$} and \textbf{H}\textbf{$_{ct}$}.}

\vspace{1mm}
\noindent \textbf{DR$_2$.}  \textit{Consider co-visualizing time across space to improve task accuracy and user confidence.} Participants were significantly more accurate with \visLabel{EmbeddedV+Xray} compared to both \visLabel{TemporalJX} and \visLabel{TemporalJX+Xray}. This result could indicate that the ability to co-visualize multiple time instances may have been a key factor that led to increased task correctness.
\highlight{Participants also showed slightly higher confidence with \visLabel{EmbeddedV+Xray} than with \visLabel{TemporalJX}}. In the post-study questionnaire, some participants reported that they felt effective and confident throughout the study because they could ``see everything in one go" and ``get the full picture of the data'', that is, they could simultaneously see and compare time instances across contiguous building regions.
\highlight{This confirms \textbf{H}\textbf{$_{err}$} and partially supports \textbf{H}\textbf{$_{conf}$}.}

In contrast, participants stated difficulties and their mitigation strategies when using either \visLabel{TemporalJX} or \visLabel{TemporalJX+Xray}. When using \visLabel{TemporalJX}, some participants mentioned that their strategy to reduce task difficulty involved positioning themselves in vantage points where they could maintain a line of sight of multiple facades simultaneously. This behavior is somewhat portrayed in some density layouts, where bright red areas appear strategically positioned \textit{e.g.} at building corners (see \textcolor{black}{Fig. \ref{fig:resultsDensity}--\textit{c}}). Nonetheless, they also noted that this strategy could become ineffective in constrained areas with a high degree of curvature, resulting in a restricted field of view where peripheral vision cannot be properly leveraged to inspect multiple facades of the building (see \textcolor{black}{Fig. \ref{fig:resultsDensity}--\textit{f}}). When using \visLabel{TemporalJX+Xray}, some participants mentioned their strategy to lower task difficulty as progressively scanning data on each facade, comparing the data with the currently-selected task answer, and updating the answer as needed.

\vspace{1mm}
\noindent \textbf{DR$_3$.}  \textit{Favor co-temporal x-ray integration.}
The results are promising, as our proposed \visLabel{EmbeddedV+Xray} design consistently yielded better results than the other visualization conditions. This technique appears to effectively combine the strengths of the two design strategies: \textit{i}) co-visualization of time steps, which contributes to increased accuracy, and \textit{ii}) x-ray-based de-occlusion, which enables faster responses.

\vspace{1mm}
\noindent \textbf{DR$_4$.}  \textit{Design to maximize data visibility in time and space.} Our study results indicate that participants consistently sought to maximize data visibility (spatially and temporally) by adopting strategies such as selecting vantage spatial locations or using task answers to track previously viewed temporal states. These behaviors underscore the need to minimize both physical and mental load, and highlight a key design consideration for future spatiotemporal 3D urban visualizations, in both immersive and conventional spaces.

\section{Conclusion, Limitations \& Future Work}
\label{sec:conclusion}
In this paper, we introduced an immersive lens visualization designed to facilitate spatiotemporal 3D urban data exploration and analysis. Through the combination of conformal mapping and view-dependent de-occlusion, our approach enables the seamless integration of spatial and temporal information within a unified space, promoting rapid space-time association and ultimately the discovery, access, and interpretation of spatiotemporal urban patterns.
A controlled evaluation validated the usage and usefulness of our immersive lens visualization, providing empirical evidence that both shape-adaptive embedding and x-ray de-occlusion design strategies benefit task efficiency, physical effort, and user confidence. 
Based on our findings, we distilled design recommendations intended to inform future research in visualization, urban, and related domains.

\vspace{1mm}
\noindent \highlight{\textbf{Limitations.}} We acknowledge \removetext{some} limitations of our work.
First, \highlight{our current prototype implementation of the conformal mapping algorithm operates as an \textit{offline} pre-processing step, \textit{i.e.}, the resulting static layout cannot be dynamically recomputed. 
As described in our supplemental material, this limitation arises from the fact that we use a Matlab implementation (the only implementation to the best of our knowledge) of the Schwarz–Christoffel (SC) algorithm.
%
%
This scenario was sufficient to perform our user study.
We leave the problem of re-implementing the SC algorithm and, therefore, making the conformal mapping computation fully interactive to future work.}
Second, our algorithm inherits limitations from the underlying (SC) formulation, which is known to suffer from numerical instability \highlight{(such as the crowding phenomenon) especially in domains with elongated or highly irregular geometries (\textit{e.g.}, building footprints with very acute angles or deeply concave indentations) may cause \removetext{disproportionate compression of prevertices in the unit disk domain, resulting in} distortion or imprecise ribbon spacing in the layered layout.} 
Despite attempting to mitigate these effects through partitioning and graph-based stitching, the fidelity of the \textit{local} conformal layouts remains sensitive to underlying geometric and numerical instabilities.
\highlight{Third, the scope of our user study was intentionally constrained for feasibility. Our findings are based on a single temporal attribute, a limited set of spatiotemporal urban tasks, and a small number of temporal and spatial instances. Consequently, more studies may be necessary to assess both the generalizability of our visualization to other urban models, spatiotemporal attributes, and tasks, and its scalability to finer temporal or spatial granularities.} These investigations might reveal alternative outcomes, user behaviors, and strategies, ultimately contributing to more reusable and robust design guidelines.

\vspace{1mm}
\noindent \highlight{\textbf{Future work.}} 
A promising research direction is to investigate glyph-based encoding designs, possibly employing layering strategies to portray multiple temporal attributes within a single lens image (\textit{e.g.} wind, temperature, and solar exposure to examine how facade comfort evolves during summer days) since prior work has shown that combining colormaps with layered glyphs can effectively enable the co-visualization of multiple attributes on smooth surfaces \cite{rocha2016}.
Additionally, although this paper primarily investigated spatially-based visual encodings, going forward we plan on exploring spatiotemporal thematic representations within our shape-adaptive lens. 
While both thematic and spatial mappings aim to visually convey data attributes, they serve distinct analytical purposes. Spatial representation, \textit{e.g.}, color mapping, promotes realism and spatial reasoning by preserving the geometric and topological fidelity of the data. In contrast, thematic mappings center on higher-level analytical reasoning by abstracting the spatial dimension to bring into focus attribute-driven patterns, such as correlations, trends, outliers, and temporal cycles. Beyond that, while 3D spaces introduce perceptual shortcomings (\textit{e.g.}, occlusion and perspective distortion), that can undermine the legibility of abstract visualizations, prior studies have shown that under certain conditions (\textit{e.g.} close-range camera views and smooth or planar surfaces) spatially-embedded abstract encodings can yield accurate results, sometimes even outperforming familiar spatial encodings such as color mapping \cite{mota2022comparison}. Thus, bridging spatial and abstract paradigms appears to be a promising direction for future research on conformal, spatially-embedded lens visualizations.

\section{Acknowledgements}

We thank the reviewers for their valuable feedback and acknowledge the support of the Natural Sciences and Engineering Research Council of Canada [RGPIN-2025-04694, RGPIN-2019-05320, CGSD3-534866-2019], the U.S. National Science Foundation [Awards \#2320261,
\#2330565, \#2411223], and the Brazilian National Council for Scientific and Technological Development CNPq [311425/2023-2].

\bibliographystyle{abbrv-doi-hyperref}

\bibliography{template}

\end{document}